\documentclass[twocolumn,aps,prb,superscriptaddress,amsmath,amssymb,showpacs]{revtex4}
\usepackage{graphicx}

\begin{document}

\title{Evidence for a very low-lying $S = 9$ excited state of the $S = 10$ single molecule magnet Mn$_{12}-$acetate}

\author{R. S. Edwards}
\affiliation{Department of Physics, University of Florida,
Gainesville, FL 32611,USA}
\author{S. Hill}
\email[corresponding author, Email:]{hill@phys.ufl.edu}
\affiliation{Department of Physics, University of Florida,
Gainesville, FL 32611,USA}
\author{S. Maccagnano}
\affiliation{Department of Physics, University of Florida,
Gainesville, FL 32611,USA}
\author{N. S. Dalal}
\affiliation{Department of Chemistry and National High Magnetic
Field Laboratory, Tallahassee, FL 32310, USA}
\author{J. M. North}
\affiliation{Department of Chemistry and National High Magnetic
Field Laboratory, Tallahassee, FL 32310, USA}

\date{\today}

\begin{abstract}
We present a detailed investigation of the temperature and
frequency dependence of the anomalous EPR transitions first
observed in Mn$_{12}$-acetate by Hill {\em et al}. [Phys. Rev.
Lett. {\bf 80}, 2453 (1998)]. The most dominant of these
transitions manifest themselves as an extra series of EPR
absorption peaks for spectra obtained with the DC field applied
within the hard magnetic plane of a single crystal sample. Recent
studies by Amigo {\em et al}. [Phys. Rev. B {\bf 65}, 172403
(2002)] have attributed these extra peaks to a strain induced
transverse quadratic anisotropy which gives rise to distinct
Mn$_{12}$-acetate species, each having a distinct EPR spectrum; on
the basis of these measurements, it has been suggested that this
transverse anisotropy is responsible for the tunneling in
Mn$_{12}$-acetate. Our temperature and frequency dependent
measurements demonstrate unambiguously that these anomalous EPR
absorptions vanish as the temperature tends to zero, thereby
indicating that they correspond to transitions from an excited
state of the molecule. We argue that this low lying excited state
corresponds to an $S = 9$ multiplet having very similar zero-field
crystal parameters to the $S = 10$ state, and lying only about
$10-15$~$k_B$ above the $S = 10$ (M$_S = \pm 9$) ground state.
These findings also compare favorably with available neutron
scattering data.
\end{abstract}


\maketitle

\noindent{{\bf I. Introduction}}

\smallskip

Since it was first suggested that macroscopic quantum tunneling of
magnetic moments from one easy axis to another should be
observable in ensembles of sufficiently tiny magnetic
particles,\cite{Bean59} the quantum properties of crystals
containing identical magnetic molecules with relatively large
magnetic moments (so far up to $26\mu_B$) $-$ so called single
molecule magnets (SMMs)\cite{MRS00,Angewandte03} $-$ have become
the subject of intense research activity. The main feature of
magnetic quantum tunneling (MQT) is the following: below a
characteristic temperature, called the blocking temperature T$_B$,
the magnetic viscosity (relaxation time for magnetization
reversal) stays finite and temperature independent, indicating
that the moments are able to short-circuit the classical thermally
activated relaxation
channel.\cite{Bean59,MRS00,Angewandte03,TejadaBook} SMMs offer a
number of advantages over other types of magnetic nanostructures.
Most importantly, they are monodisperse. Consequently, they enable
fundamental studies of properties intrinsic to magnetic
nanostructures that have previously been inaccessible. For
example, the observation of well defined quantum jumps in the low
temperature (T~$<$~T$_B$) hysteresis loops of
[Mn$_{12}$O$_{12}$(CH$_3$COO)$_{16}$(H$_2$O)$_4$]$\cdot$2CH$_3$COOH$\cdot4$H$_2$O
(Mn$_{12}$-ac) crystals provided evidence for resonant
MQT.\cite{MRS00,Angewandte03,TejadaBook,Friedman96,Thomas96} Since
this discovery, Mn$_{12}$-ac has become the most widely studied
SMM.\cite{MRS00,Angewandte03,TejadaBook,Friedman96,Thomas96,Hernandez97,delBarcoEPL02,Hennion97,MirebeauPRL99,Barra97,HillPRL98,HillPRB02,KPark02a,KPark02b,HillPRL03a}
The resonant nature of the MQT has enabled very precise studies of
the relaxation dynamics,\cite{Hernandez97,delBarcoEPL02} while
neutron\cite{Hennion97,MirebeauPRL99} and electron paramagnetic
resonance
(EPR)\cite{Barra97,HillPRL98,HillPRB02,KPark02a,KPark02b,HillPRL03a}
experiments have provided fairly detailed spectroscopic
information concerning the low-lying quantum energy levels of
Mn$_{12}$-ac. In spite of these extensive investigations, the
precise mechanism responsible for the tunneling has remained
elusive. An in-depth understanding of MQT is important not only
from a fundamental point of view: for example, there has recently
been considerable speculation concerning the possible use of SMMs
for quantum information processing.\cite{LossNature01}

The Mn$_{12}$-ac molecule consists of four Mn$^{4+}$ ions, each
with spin $S = \frac{3}{2}$, surrounded by eight Mn$^{3+}$ ions
with spin $S = 2$.\cite{MRS00,Angewandte03} The clusters
crystallize into a tetragonal lattice, with each molecule having
approximate $S_4$ site symmetry; the orbital moment is quenched,
and a Jahn-Teller distortion produces a strong axial anisotropy. A
simplified treatment of the magnetic interactions within the
molecule has been developed\cite{SessoliJACS93} wherein four
strongly antiferromagnetically coupled Mn$^{3+}-$Mn$^{4+}$ dimers,
each with spin $S = 2 - \frac{3}{2} = \frac{1}{2}$, couple via an
effective ferromagnetic interaction to the remaining four $S = 2$
Mn$^{3+}$ ions, giving a total spin $S = 10$. Within this
simplified 8-spin scheme, the weaker couplings between the four
spin$-\frac{1}{2}$ dimers and the four spin$-2$ Mn$^{3+}$ ions
largely determine the low energy excitations within the
molecule.\cite{Katsnelson99}

To lowest order, the magnetic energy levels of a rigid spin $S =
10$ system can be described by the effective spin Hamiltonian:

\bigskip

\noindent{ \hfill  \hfill  \hfill  \hfill  \hfill $\hat H = D\hat
S_z^2 + \mu _B \vec B \cdot
\mathord{\buildrel{\lower3pt\hbox{$\scriptscriptstyle\leftrightarrow$}}
\over g}  \cdot \hat S + \hat H', \hfill\hfill\hfill\hfill\hfill
(1)$}

\bigskip

\noindent{where $D$ $(< 0)$ is the uniaxial anisotropy constant,
the second term represents the Zeeman interaction with an applied
field $\vec{B}$
($\mathord{\buildrel{\lower3pt\hbox{$\scriptscriptstyle\leftrightarrow$}}
\over g}$ is the Land$\acute{e}$ g tensor), and $\hat{H}^\prime$
includes higher order terms in the crystal field ($\hat{O}^0_4$,
$\hat{O}^2_2$, $\hat{O}^2_4$, $\hat{O}^4_4$, {\em etc.}
\cite{Hennion97,MirebeauPRL99,Barra97,HillPRL98}), as well as
environmental couplings such as intermolecular dipolar and
exchange interactions.\cite{HillPRB02,KPark02a,KPark02b} This
predominantly Ising-type anisotropy is responsible for the energy
barrier to magnetization reversal and the resulting magnetic
bistability - factors which lead to magnetic hysteresis at
sufficiently low temperatures.\cite{MRS00,Angewandte03,TejadaBook}
MQT in zero-field is caused by interactions in $\hat{H}^\prime$
which lower the symmetry of the molecule from strictly axial,
thereby mixing otherwise degenerate pure "spin-up" and "spin-down"
states. Tunnel rates depend on the degree of symmetry breaking,
which is something which can be determined very precisely via
single crystal EPR
measurements.\cite{Barra97,HillPRL98,HillPRB02,KPark02a,KPark02b,HillPRL03a}
The only transverse crystal field term (to fourth order) in
$\hat{H}^\prime$ allowed by the strict $S_4$ symmetry of the
molecule is $\hat{O}^4_4$. While such an interaction clearly
exists,\cite{MirebeauPRL99,Barra97,HillPRL03a} it cannot account
for many aspects of the low temperature magnetic
relaxation.\cite{delBarcoEPL02,FriedmanRC98} For this reason,
recent theoretical and experimental attention has focused on a
quadratic transverse anisotropy induced by
disorder.\cite{delBarcoEPL02,HillPRB02,KPark02a,KPark02b,HillPRL03a,ChudnPRL01,GaraninPRB02,AmigoPRB02,ParksPRB01,MertesPRL02,CorniaPRL02}
One particular proposal considers long range strains caused by
dislocations.\cite{ChudnPRL01,GaraninPRB02} However, a more likely
scenario involves acetate ligand disorder which would give rise to
distinct Mn$_{12}$-ac variants with local site-symmetry lower than
$S_4$.\cite{delBarcoEPL02,HillPRB02,HillPRL03a,CorniaPRL02}
Indeed, our recent angle dependent EPR studies show clear evidence
for a quadratic anisotropy ($\hat{O}^2_4$) which can be attributed
to a discrete disorder associated with the ligand
molecules.\cite{HillPRL03a} Furthermore, recent magnetic
measurements have shown that a narrow distribution of tunnel
splittings exists,\cite{delBarcoEPL02} a fact which is again
consistent with the ligand disorder scenario.\cite{CorniaPRL02}}

In spite of growing acceptance for the role of disorder in the MQT
phenomenon, several outstanding questions remain; not least, the
absence of a clear parity effect associated with the resonant MQT
steps observed in hysteresis experiments.\cite{FriedmanRC98}
Meanwhile, several recent works have drawn attention to possible
inadequacies of the single spin ($S = 10$) description of
Mn$_{12}$-ac.\cite{Hennion97,Katsnelson99,AcheyPRB01,AcheySSC02,YamamotoRPL02}
For example: a low lying magnetic excitation has been observed in
neutron scattering experiments, which cannot be explained within
the single spin picture;\cite{Hennion97} NMR experiments have
demonstrated that the paramagnetic spin density is delocalized
over the entire molecule;\cite{AcheyPRB01,AcheySSC02} and recent
calculations\cite{Katsnelson99} have shown that interactions
between the $S = 10$ multiplet and low lying excited multiplets
($S < 10$) are necessary in order to explain the significant
quartic terms in Eq. (1) which are found for
Mn$_{12}$-ac.\cite{MirebeauPRL99,Barra97,HillPRL03a} In this
paper, we present high-frequency single crystal EPR data which
provide convincing evidence for a low-lying $S = 9$ state. We
compare these findings with available neutron scattering
data,\cite{Hennion97} and comment on the possible consequences of
these findings.

\bigskip

\noindent{{\bf II. Experimental}}

\smallskip

The $(2S+1)-$fold quantum energy level structure associated with a
large molecular spin necessitates spectroscopies spanning a wide
frequency range. Furthermore, large zero-field level splittings,
due to the significant crystalline anisotropy (large $D$) and
large total spin $S$, demand the use of frequencies and magnetic
fields considerably higher (50~GHz to 1~THz, and up to 10~tesla
respectively) than those typically used by the majority of EPR
spectroscopists. The high degree of sensitivity required for
single crystal measurements is achieved using a resonant cavity
perturbation technique in combination with a broad-band
Millimeter-wave Vector Network Analyzer (MVNA) exhibiting an
exceptionally good signal-to-noise ratio; a detailed description
of this instrumentation can be found in
ref.~[\onlinecite{MolaRSI}]. The MVNA is a phase sensitive, fully
sweepable (8 to 350~GHz), superheterodyne source/detection system.
Several sample probes couple the network analyzer to a range of
high sensitivity cavities ($Q-$factors of up to 25,000) situated
within the bore of a superconducting magnet. The MVNA/cavity
combination has been shown to exhibit a sensitivity of at least
$10^9$~spins$\cdot$G$^{-1}\cdot$s$^{-1}$, which is comparable with
the best narrow-band EPR spectrometers. This, coupled with newly
acquired sources and a split-pair magnet, allow single crystal
measurements at any frequency in the range from 8 to 250~GHz, at
temperatures down to 1.2~K $(\pm 0.01$~K), and for any geometrical
combination of DC and AC field orientations.

The advantages of a narrow band cavity perturbation technique, and
the ability to study single crystals, have recently been discussed
in ref.~[\onlinecite{HillPRB02}]. In particular, such a scheme
enables faithful extraction of the intrinsic EPR lineshapes (both
the real and imaginary components), free from instrumental
artifacts. Consequently, any raw data displayed in this paper
constitutes pure absorption. Single Mn$_{12}$-ac crystals were
grown using literature methods.\cite{Lis80} All measurements were
performed in the standard EPR configuration with the AC excitation
field transverse to the DC field. The two samples (A and B) used
in this study were needle shaped, having approximate dimensions
$1\times 0.1\times 0.1$~mm$^3$. Orientation of the crystals was
relatively straightforward due to the shape of the samples, with
the needle axis defining the easy axis. All of the presented data
were obtained with the DC magnetic field aligned within the
samples' hard magnetic $(x,y)$ plane (easy axis data are presented
in ref.~[\onlinecite{HillPRB02}]). Low field data $(< 7$~T) were
obtained using a superconducting solenoid, while higher field data
were obtained in the resistive magnets at the National High
Magnetic Field Laboratory (NHMFL).

\bigskip

\noindent{{\bf III. Background}}

\smallskip

In our earlier single crystal investigations of Mn$_{12}$-ac, we
first pointed out that EPR spectra obtained with the field
perpendicular to the easy-axis revealed a number of anomalous EPR
transitions;\cite{HillPRL98} these transitions were labeled
$\beta$ and $\gamma$ in ref.~[\onlinecite{HillPRL98}], as opposed
to the $\alpha$-resonances which nicely fit the accepted $S = 10$
Hamiltonian for Mn$_{12}$-ac (see Fig.~1 and Table~1, which adopts
this same labeling scheme). These earlier studies were conducted
at moderate fields $(B < 11$~T), and at relatively low frequencies
compared to the powder studies by Barra {\em et al}.\cite{Barra97}
In the high-field limit, one expects 20 transitions within the $2S
+ 1$ $(S = 10)$ multiplet; the $\alpha$-resonances in
ref.~[\onlinecite{HillPRL98}] and Fig.~1 comprise half of this
total (corresponding to transitions from M$_S =$~even-to-odd
integers, see Table~1). One should eventually expect to see an
additional 10 EPR peaks, in between the $\alpha$-resonances
(corresponding to transitions from M$_S =$~odd-to-even integers),
but only in the high-field/frequency limit; Fig.~2 illustrates the
origin of these transitions. We previously tentatively ascribed
the $\beta$-resonances to the expected high-field resonances.
However, the M$_S =$~odd-to-even transitions should become EPR
silent below about 95~GHz (due to avoided level crossings, see
Fig.~2b), yet they persist down to at least 45~GHz. Indeed, all
attempts to fit the $\beta$-resonances to the accepted
Mn$_{12}$-ac Hamiltonian (Eq.~1) have failed quite spectacularly.
We should point out that we have since carried out single crystal
studies on simpler $S = \frac{9}{2}$ Mn$_4$ SMMs (corresponding to
the core of the Mn$_{12}$ SMM), with the same $S_4$ symmetry,
which behave absolutely as expected according to
Eq.~1.\cite{SB4Poly02}

More recently, several other groups have reported extra EPR peaks
for both single crystal studies with the field perpendicular to
the easy axis, and from powder measurements. Amigo {\em et
al}.\cite{AmigoPRB02} ascribe the extra peaks to a strain or
disorder induced transverse anisotropy (caused by dislocations in
the sample, see ref.~[\onlinecite{ChudnPRL01,GaraninPRB02}]),
which gives rise to different species of molecule within the
crystal. In this model, the extra peaks arise due to the fact that
one expects equal numbers of molecules having their hard axes
along a particular direction in the hard plane, and at $90^\circ$
to this direction. Matters are complicated further due to the fact
that their model includes a rather broad distribution of
transverse anisotropies. Thus, it is unlikely that one would
observe distinct splittings in such a case, but rather a
broadening of the resonance linewidths, with an EPR lineshape that
reflects the distribution. Cornia {\em et al}. have also recently
published data which show a splitting of the $\alpha$-resonances
(the $\beta$ resonances are also seen in their 95~GHz
spectra).\cite{CorniaPRL02} However, this splitting is more
subtle, since it is only clearly observed using field modulation,
or by taking derivatives of absorption spectra.\cite{HillPRL03a}
We also see this effect in the $\alpha$-resonances (this is the
subject of a separate publication\cite{HillPRL03a}). Cornia {\em
et al}. propose a mechanism for this splitting involving disorder
associated with the acetic acids of crystallization. Like Amigo's
model, this ligand disorder induces a transverse anisotropy which
varies from one molecule to the next.

We believe that the EPR splittings observed by Amigo {\em et
al}.,\cite{AmigoPRB02} and by Cornia {\em et
al}.,\cite{CorniaPRL02} have quite different origins. The latter
almost certainly are associated with ligand disorder, as we have
recently confirmed,\cite{HillPRL03a} whereas the distinct
absorptions observed by Amigo {\em et al}. correspond to the same
$\beta$-resonances that we observed in our original single crystal
investigations.\cite{HillPRL98} Although Amigo {\em et al}. claim
that the anomalous peaks are only observed upon stressing their
samples, there is clear evidence for such resonances even in their
unstressed samples (see Fig.~2 in ref.~[\onlinecite{AmigoPRB02}]].
Furthermore, we have seen these extra peaks in all of the samples
that we have studied, as have Barra {\em et al}.\cite{Barra97} and
Cornia {\em et al}.\cite{CorniaPRL02} There do, however, appear to
be noticeable differences in the positions and intensities of
anomalous EPR peaks obtained by different groups. Since transverse
EPR spectra are notoriously sensitive to field
alignment,\cite{HillPRL03a} minor differences in sample
orientation may easily explain these differences, as could solvent
loss from the samples.

\bigskip

\noindent{{\bf IV. Results and discussion}}

\smallskip

Figure~1 shows data obtained for sample A at four different
frequencies from 44~GHz up to 111~GHz; the temperature is 10~K for
each sweep. The dips in absorption are due to EPR transitions, and
these have been labeled according to the original scheme in
ref.~[\onlinecite{HillPRL98}]. The $\alpha$-resonances correspond
to transitions within the Zeeman split energy levels which, in
zero-field, correspond to M$_S = \pm m$ states $(m =$~integer, and
$0 \leq m \leq S$) as depicted in Fig.~2a; in this representation,
the quantization axis is defined by the uniaxial crystal field
tensor, and is along $z$. In the high-field limit, the
quantization axis points along the applied field vector; in this
limit, the ten $\alpha$-resonances correspond to transitions from
M$_S =$~even-to-odd $m$, {\em e.g}. M$_S = -10$ to $-9$ (again,
this situation is illustrated in Fig.~2a). Because the
$\alpha$-resonances originate from pairs of levels which are
(approximately) degenerate in zero field, one expects the
resonance frequencies, when plotted against field, to tend to zero
as the field tends to zero; this can be seen in Fig.~2b.

In the high field limit, one expects ten additional resonances
corresponding to M$_S =$~odd-to-even $m$, {\em e.g}. M$_S = -9$ to
$-8$. Indeed, we do see these transitions at high fields (see
below). However, if one follows these transitions to the
zero-field limit, it is clear that the resonance frequency should
go through a minimum, tending to one of the zero-field splittings
in this limit; this is illustrated in Fig.~2a and by the dashed
curves in Fig.~2b. Consequently, such EPR transitions should
become "EPR silent" below some characteristic cut-off frequency
determined by the point of closest approach for the two levels
involved in the transition. The dominant interaction responsible
for this level repulsion is simply the Zeeman term in Eq.~1 and
is, therefore, more-or-less insensitive to the other transverse
terms in Eq.~1. For Mn$_{12}$-ac, this cut-off frequency is about
95~GHz for the three highest field odd-to-even $m$ transitions
(see Fig.~2b, which uses realistic parameters for
Mn$_{12}$-ac).\cite{MirebeauPRL99,HillPRL03a} For this reason,
below 95~GHz, the $\beta$-resonances observed in Fig.~1 cannot be
attributed to the high-field odd-to-even $m$ transitions. However,
it is noticeable that the strength of the $\beta$-resonances
increases dramatically between the 77.4~GHz trace and the
111.1~GHz trace, {\em i.e}. above 95~GHz, there is a pronounced
increase in the strength of the $\beta$-resonance. We shall
comment more on this below.

Next, we turn to the temperature dependence of the data obtained
in Fig.~1, with particular attention being paid to the highest
field $\beta$-resonance ($\beta10$), and a much weaker
$\gamma$-resonance. Fig.~3 shows data for three frequencies below
the cut-off of 95~GHz. In all of these traces, the $\beta10$
resonance diminishes in intensity as T~$\rightarrow0$, becoming
more-or-less invisible at the lowest temperature. This fact proves
beyond any doubt that the $\beta10$ resonance originates from an
excited state of the Mn$_{12}$-ac molecule, thereby ruling out any
likelihood that it corresponds to a different Mn$_{12}$-ac
species, as proposed by Amigo {\em et al}.\cite{AmigoPRB02} We
will discuss the possible origin of this excited state further
below. Meanwhile, we note that the weak $\gamma$-resonance
exhibits a very similar temperature dependence to the $\alpha10$
resonance, which corresponds to the transition from the accepted
$S = 10$ ground state of the molecule, {\em i.e}. the $\gamma$ and
$\alpha10$ peaks continue to grow as T~$\rightarrow0$. Therefore,
there is every likelihood that the $\gamma$ peak does signify a
small minority ($\sim 1$~to~$2\%$) of molecules having a different
ground state from the majority. We speculate that this peak may
correspond to the well known faster relaxing species of
Mn$_{12}$-ac, which has one of its Jahn-Teller axes tilted with
respect to the $z$-axis of the molecule.\cite{Sun99,Aubin01} This
peak has only been observed for some samples, a fact which is also
consistent with the faster relaxing species. The faster relaxing
Mn$_{12}$-ac Jahn-Teller isomer is believed to have an $S = 9$
ground state, a reduced axial anisotropy constant ($D$-value) and
an appreciable transverse quadratic anisotropy.\cite{ChristouPC}
All of these factors can explain the fact that the
$\gamma$-resonance is observed at lower fields than the $\alpha
10$ resonance. However, the shift appears to be rather large,
implying a $\sim 30\%$ reduction in $D$ over the majority species;
further studies on pure samples of the faster relaxing species are
planned in the near future in order to resolve this issue. A
similar weak lower lying peak has been observed in neutron
scattering experiments where, again, the authors attribute this to
the faster relaxing Mn$_{12}$-ac species.\cite{MirebeauPRL99} It
is interesting to note that the ratios of the frequencies of the
presumed fast and slow relaxing peaks observed here, and from
neutron scattering are quite similar,\cite{MirebeauPRL99} as are
the relative intensities.

Figure~4 shows the temperature dependence of the 111.1~GHz data
from Fig.~2. At this frequency, which is above the cut-off for
odd-to-even $m$ high-field transitions, the $\beta10$ resonance
shows a marked increase in intensity over the lower frequency
data. Indeed, it exceeds the intensity of the $\alpha10$ resonance
at temperatures above about 4.2~K. Furthermore, $\beta10$ is
clearly visible at the lowest temperatures investigated. A similar
trend may be noted for $\beta9$. We suspect that this pronounced
enhancement of $\beta10$ (and $\beta9$) is caused by the switching
on of the anticipated high-field M$_S = -9$~to $-8$ ($-7$ to $-6$)
transition within the $S = 10$ state of Mn$_{12}$-ac (see Fig.~2
and Table~1); the fact that it is difficult to distinguish these
transitions from the anomalous low frequency transitions will be
discussed further below. EPR spectra obtained with the field
applied parallel to the easy ($z$-) axis show no additional
anomalous peaks,\cite{HillPRB02} {\em i.e}. no EPR absorptions
which cannot be explained within the framework of the $S = 10$
picture. Easy axis data obtained over a wider frequency range (40
to 200~GHz) compared to our earlier studies
(ref.~[\onlinecite{HillPRL98}]) have recently been published in
ref.~[\onlinecite{HillPRB02}].

Recent hard axis measurements for a second sample (B), covering a
wider frequency range compared to our original
investigations,\cite{HillPRL98} have enabled a complete
determination of the Hamiltonian parameters for Mn$_{12}$-ac up to
fourth order.\cite{HillPRL03a} Further details of our fitting
procedure are published elsewhere.\cite{HillPRL03a} Fig.~5 shows
the results of such a fit, along with the obtained parameters. As
can be seen, the solid curves $(S = 10$ picture) lie beautifully
on the $\alpha$-resonance data (open circles). However, as already
discussed, the $\beta$-resonances (solid squares) only match the
$S = 10$ curves at the highest frequencies investigated; below
95~GHz, a dramatic departure from the expected behavior can be
seen. It is impossible to force the $S = 10$ fits through these
low frequency $\beta$-resonance data points, since it is apparent
that they belong to curves which approach zero frequency as the
field tends to zero; for an $S = 10$ system, the high-field
odd-to-even $m$ transitions tend to finite frequency offsets in
the low field limit (see Fig.~2). We therefore speculate that, at
low frequencies, the $\beta$-resonances correspond to transitions
within an excited state of Mn$_{12}$-ac. For comparison, Fig.~5
includes curves corresponding to the $S = 9$ state, which were
generated using precisely the same parameters as obtained from the
$S = 10$ fit. For an odd total spin state, the low field limiting
behavior of odd-to-even $m$ and even-to-odd $m$ transitions is the
reverse of that for an even total spin state. Consequently, one
{\em does expect} the frequency of the M$_S = -9$ to $-8$
transition to go to zero in the low field limit for $S = 9$.
Although not perfect, the low frequency $\beta$-resonance data lie
quite close to the $S = 9$ curves. At higher frequencies, the
$\beta$-resonances lie somewhat in between the $S = 9$ and $S =
10$ curves while, in the high frequency limit, they lie nicely on
the $S = 10$ curves.

Having noted the possible connection between the
$\beta$-resonances and an $S = 9$ state, one has to ask whether
such an assignment is consistent with other published results.
Based on the limited temperature dependence of $\beta10$ (Fig.~3),
we can estimate the approximate activation energy to the state
from which $\beta10$ is excited, {\em i.e}. the presumed $S = 9$
state. This analysis is complicated somewhat by the broad
asymmetric $\alpha 10$ resonance, and due to the fact that the
$\beta10$ resonance saturates at temperatures as low as $4-6$ K.
Nevertheless, a crude analysis yields a very low activation energy
of 10-15 $k_B$. Incredibly, this implies that the supposed $S = 9$
(M$_S = \pm 9$) state lies very close to the first excited state
of the $S = 10$ multiplet! Even looking by eye at the data in
Fig.~3, one can see that the temperature dependence of $\beta10$
and $\alpha9$ are quite similar, so it is clear that the
activation energy to the $S = 9$ state is of a similar order to
the low lying $S = 10$ levels. An energy of $10$~$k_B$ corresponds
precisely to 0.2~THz, where a very strong finite $Q$
(1.18~$\AA^{-1}$) mode has been observed from neutron scattering
experiments.\cite{Hennion97} The momentum dependence of the
intensity of this mode indicates that it has a magnetic origin
(its intensity tends to zero as $Q\rightarrow0$), suggesting spins
that are coupled on a local scale. Indeed, from an analysis of the
neutron data, one can conclude that this mode may be related to
the antiferromagnetic coupling within a Mn$^{3+}-$Mn$^{4+}$ dimer
within the framework of the Mn$_{12}$ molecule.\cite{Hennion97}

Recent theoretical studies of the spin excitations within
Mn$_{12}$-ac treat the molecule as a system of four strongly
antiferromagnetically coupled Mn$^{3+}-$Mn$^{4+}$ dimers, each
with spin $S = 2 - \frac{3}{2} = \frac{1}{2}$ which, in the ground
state, are coupled via an effective ferromagnetic interaction to
the four remaining Mn$^{3+}$ ions, each having $S =
2$.\cite{SessoliJACS93,Katsnelson99,YamamotoRPL02} Within this
simplified scheme, the weaker couplings between the four
spin$-\frac{1}{2}$ dimers and the four spin$-2$ Mn$^{3+}$ ions
largely determine the low energy excitations within the molecule.
While these calculations do not reproduce the 0.2~THz neutron
mode,\cite{Hennion97} this 8 spin model provides a basis for
comparison with the present data. In particular, one could imagine
an excitation within a spin$-\frac{1}{2}$ dimer leading to a
reversal of this moment, and to an overall $S = 9$ state for the
molecule. It was previously speculated that the 0.2~THz mode
originates from dissipative interactions between the Mn$_{12}$
clusters and their environment.\cite{Katsnelson99} The present
work seems to indicate that this is not the case, and that the
0.2~THz mode corresponds to a distinct finite $Q$ excitation of
the molecule to a well defined (probably $S = 9)$ state. We hope
that these findings will stimulate further theoretical work on
this problem.

While our measurements do not provide definitive proof that the
low lying excitation within the molecule corresponds to an $S = 9$
state, such an explanation for the observed trends in the single
crystal EPR data is quite appealing. Furthermore, it is apparent
that the $S = 9$ and $S = 10$ data become indistinguishable in the
high frequency limit, which would explain why other researchers
have successfully simulated most aspects of their high-field
powder EPR data using an $S = 10$ model.\cite{Barra97,AmigoPRB02}
A summary of the tentative assignments of the resonances is
presented in Table~1. In order to simplify the ensuing discussion,
we shall assume that the observed excited state does, in fact,
correspond to $S = 9$. The fact that the $\beta$ resonances merge
smoothly into the $S = 10$ model at high frequencies suggests that
the parameters describing the $S = 9$ state of the molecule are
quite similar to those describing the $S = 10$ state. To a first
approximation, this is not so unrealistic since this merely
corresponds to the reversal of the moment of one of the dimers.
Consequently, one would expect excitations within the $S = 9$ and
$S = 10$ multiplets to be virtually indistinguishable for fields
parallel to the easy axis, since there is no distinction between
even-to-odd $m$ and odd-to-even $m$ transitions for this
orientation. This fact, again, appears to be consistent with our
experimental findings. Minor mismatches in the parameters could
give rise to temperature dependent line shapes and widths. In
particular, one could expect some line narrowing upon depopulating
the $S = 9$ state at low temperatures. Such behavior has been
observed, and explained very well in terms of inter-molecular
dipolar interactions.\cite{HillPRB02,KPark02a,KPark02b} Without
more precise data for the $S = 9$ state, it is not possible to say
whether this picture needs to be re-evaluated.

Due to the finite momentum separating the $S = 9$ and $S = 10$
states, it is clear that there could be no direct evidence for the
0.2~THz excitation from optical studies, {\em i.e.} no direct
optical transitions from the $S = 10$ ground state, to the
low-lying $S = 9$ state. Indeed, no such transitions ($\sim
0.2$~THz) have been reported. However, excitations within each
manifold ($S = 10$ and $S = 9$) are allowed, as appears to be the
case from these EPR investigations. Furthermore, the finite
momentum separation would tend to minimize any interaction between
these two states, thereby explaining why the $S = 10$ picture
explains so many low-temperature experimental observations with
such precision, including most EPR data. Nevertheless, even very
weak interactions between these states could have crucial
implications as far as the tunneling is concerned in Mn$_{12}$-ac.
Clearly further theoretical studies are required.

\bigskip

\noindent{{\bf V. Summary and conclusions}}

\smallskip

Detailed single crystal EPR investigations of Mn$_{12}$-ac reveal
several transitions in the hard axis spectra which do not fit the
accepted $S = 10$ picture; indeed, the deviations of these EPR
peaks from the $S = 10$ theory are quite dramatic. Temperature
dependent studies indicate that the anomalous transitions
originate from an excited state of the molecule which lies only
about $10-15$~$k_B$ above the $S = 10$, M$_S = \pm 10$ ground
state, {\em i.e}. approximately coincident with the first excited
state of the $S = 10$ multiplet. These observations conflict with
the recent suggestion by Amigo {\em et al}. that the anomalous
peaks correspond to unique disorder-induced ground states of
Mn$_{12}$-ac.\cite{AmigoPRB02}

Independent evidence for a low-lying ($\sim 10-15$~$k_B$)
excitation to something other than the standard $S = 10$ state has
been known for some time from neutron scattering
experiments.\cite{Hennion97} The present EPR investigations show,
unambiguously, that this low-lying state is very real; indeed,
both studies agree on the activation energy. Based on extensive
frequency dependent measurements, we argue that the excited state
corresponds to an $S = 9$ multiplet having very similar zero-field
crystal field parameters to the $S = 10$ state. While our
measurements do not provide definitive proof for such an $S = 9$
state, this assignment would explain why there is little evidence
for its existence in most other EPR and optical studies published
to date. The effects of this low-lying $S = 9$ state on the
low-temperature quantum properties of Mn$_{12}$-ac are not known,
and we hope these investigations will stimulate further
theoretical studies.

\bigskip

\noindent{{\bf VI. Acknowledgements}}

\smallskip

We thank Andy Kent, George Christou and David Hendrickson for
useful discussion. This work was supported by the NSF (DMR0103290,
DMR0196430 and DMR0239481); the NHMFL is supported by the State of
Florida and the NSF under DMR0084173. S. H. would like to thank
the Research Corporation for financial support.

%\bibliography{Mn12S9}

\clearpage

\begin{table}

\caption{\label{table1} Tentative assignments of the $\alpha$- and
$\beta$-resonances observed in this study, based on the high-field
representation in which M$_S$ represents the spin projection along
the applied field axis. HF and LF refer respectively to high- and
low-frequency limits.}
\begin{tabular}{cccc}
\hline Resonance & HF/LF & M$_S$ & Spin \\ \hline \hline
$\alpha10$ & HF & $-10\rightarrow-9$ & $10$ \\ $\alpha9$ & HF &
$-8\rightarrow-7$ & $10$ \\ $\alpha8$ & HF & $-6\rightarrow-5$ &
$10$ \\ $\alpha7$ & HF & $-4\rightarrow-3$ & $10$ \\ $\beta10$ &
HF & $-9\rightarrow-8$ & $10,9$ \\ $\beta9$ & HF &
$-7\rightarrow-6$ & $10,9$ \\ $\beta8$ & HF & $-5\rightarrow-4$ &
$10,9$ \\ $\beta10$ & LF & $-9\rightarrow-8$ & $9$ \\ $\beta9$ &
LF & $-7\rightarrow-6$ & $9$ \\ \hline
\end{tabular}
\end{table}

\clearpage

\noindent{{\bf Figure captions}}

\bigskip

FIG.~1. Frequency dependence of the hard axis EPR spectra for
Mn$_{12}$-ac (sample A); the temperature and the frequencies are
indicated in the figure. The $\alpha$-resonances behave as
expected for a spin $S = 10$ easy axis system. The anomalous
$\beta$-resonances increase in intensity with increasing
frequency, becoming particularly prominent at the highest
frequency; these transitions are indicated with dots ($\beta 10$)
and arrows ($\beta 9$). Assignments of the various resonances are
listed in Table~1, and Fig.~2 illustrates the possible spin $S =
10$ transitions.

\bigskip

FIG.~2. The upper panel shows the energy level diagram for
Mn$_{12}$-ac for a field applied within the hard plane; this
simulation was made using accepted Hamiltonian parameters for
Mn$_{12}$-ac ($D = -0.454$~cm$^{-1}$, $B_4^0 = -2.0 \times
10^{-5}$~cm$^{-1}$, $B^4_4 = \pm 3.0 \times
10^{-5}$~cm$^{-1}$).\cite{MirebeauPRL99,HillPRL03a} The vertical
bars illustrate the origin of the $\alpha$-resonances (even-to-odd
M$_S$ transitions), while the filled circles represent the high
frequency odd-to-even M$_S$ transitions. The quantum state labels
on the left hand side of the figure correspond to the spin
projection along $z$ (low-field limit), while the labels on the
right hand side correspond to the spin projection along the
applied field direction (high-field limit). The lower panel plots
the expected transition frequencies, as a function of magnetic
field, for the transitions labeled in the upper panel. The main
point to note is the fact that the odd-to-even M$_S$ transition
frequencies (dashed lines) go through a minimum. Therefore, these
transitions should not be observable below a cut-off frequency
which is about 95~GHz for the first few resonances in
Mn$_{12}$-ac.

\bigskip

FIG.~3. Temperature dependence of the hard axis EPR spectra
obtained at three frequencies below the hard axis cut-off
frequency (see Fig.~2 for explanation) for sample A; the
temperatures and frequencies are indicated in the figure. The main
point to note is that only the $\alpha$- and $\gamma$-resonances
persist as T~$\rightarrow0$; the intensities of the
$\beta$-resonances, on the other hand, go to zero as
T~$\rightarrow0$. Consequently, only the $\alpha$- and
$\gamma$-peaks can be attributed to ground states of Mn$_{12}$-ac
- see main text for further discussion.

\bigskip

FIG.~4. Temperature dependence of the hard axis EPR spectra
obtained above the hard axis cut-off frequency (see Fig.~2 for
explanation) for sample A; the temperature and the frequencies are
indicated in the figure. In contrast to Fig.~3, the
$\beta$-resonance shows appreciable intensity, even at the lowest
temperature while, at the highest temperature, it is by far the
strongest peak.

\bigskip

FIG.~5. Fits of the frequency dependence of hard axis spectra
obtained for sample B. Details of this analysis are published
elsewhere.\cite{HillPRL03a} The $\alpha$-resonances (open circles)
fit the $S = 10$ model (solid lines) exceptionally well. The
$\beta$-resonances (solid squares), on the other hand, only fit
the $S = 10$ picture at higher frequencies. It is not possible to
force the $S = 10$ fits through the $\beta$ resonance data points
observed below 95~GHz. However, it is found that these
low-frequency points lie quite close to the behavior expected for
$S = 9$ (dashed curves). The inset depicts the orientation of the
applied field relative to the sample, and the obtained quadratic
and quartic Hamiltonian parameters are indicated in the figure.

\bigskip

\clearpage
\begin{figure}

\includegraphics{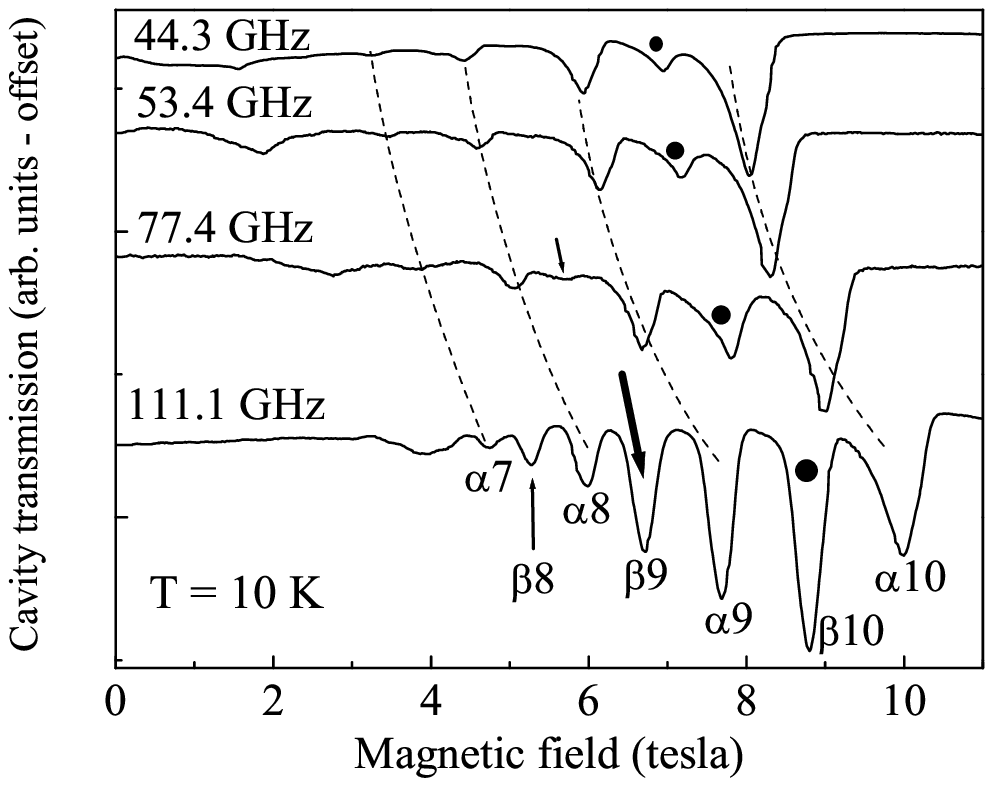}
\caption{\label{fig1} S. Hill {\em et al.}}
\end{figure}
%\clearpage

\bigskip

%\begin{turnpage}
\begin{figure}

\includegraphics{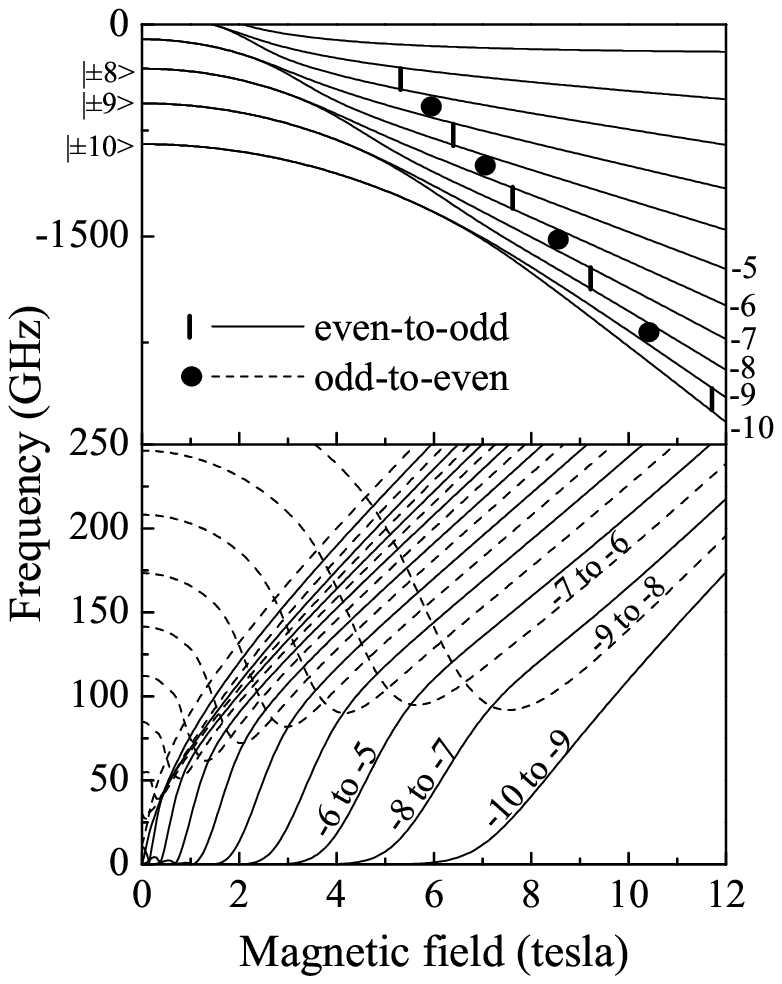}
\caption{\label{fig2} S. Hill {\em et al.}}
\end{figure}
%\clearpage
%\end{turnpage}

\bigskip

\begin{figure}

\includegraphics{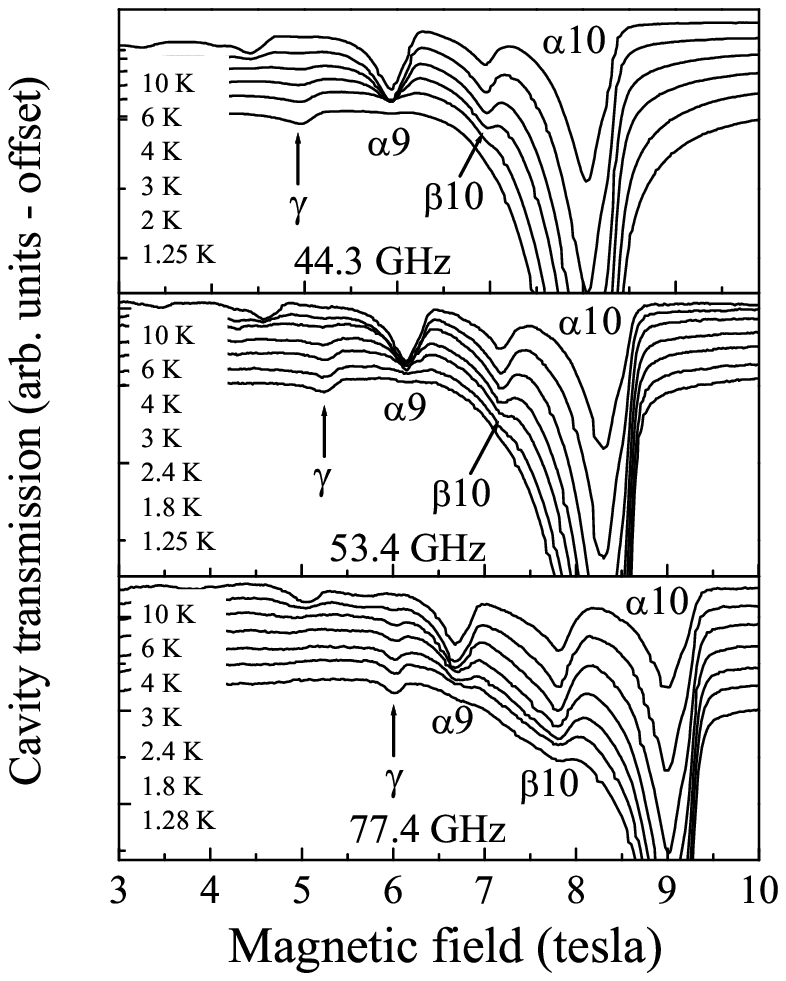}
\caption{\label{fig3} S. Hill {\em et al.}}
\end{figure}
%\clearpage

\bigskip

%\begin{turnpage}
\begin{figure}

\includegraphics{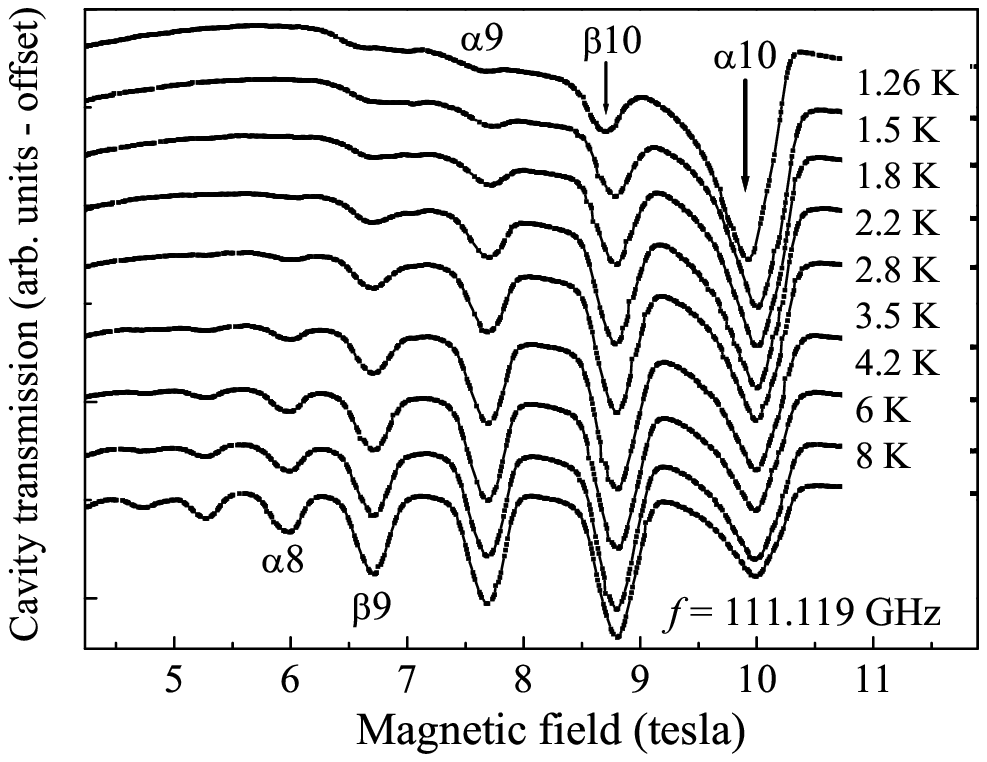}
\caption{\label{fig4} S. Hill {\em et al.}}
\end{figure}
%\clearpage
%\end{turnpage}

\bigskip

\begin{figure}

\includegraphics{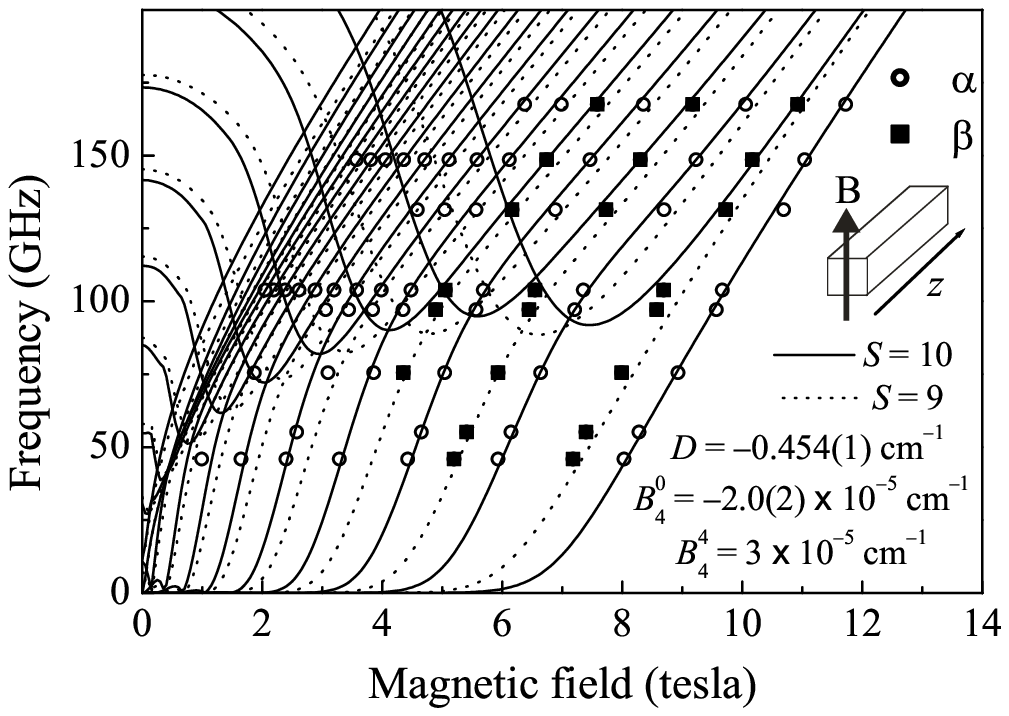}
\caption{\label{fig5} S. Hill {\em et al.}}
\end{figure}

\end{document}